\documentclass[%
 reprint,
superscriptaddress,
 amsmath,amssymb,
 aps,
 prl,
 dvipsnames,
]{revtex4-2}

\pdfoutput=1
\usepackage[utf8]{inputenc}
\usepackage[english]{babel}
\usepackage{amsmath}
\usepackage{amssymb}
\usepackage{epsfig}
\usepackage{fdsymbol}
\usepackage{tikz}

\usepackage{float}
\usepackage{graphicx}
\usepackage{dcolumn}
\usepackage{bm}

\usepackage{color}
\definecolor{mypink1}{rgb}{0.8392,0.2745,0.0353}
\definecolor{mypink2}{rgb}{0.0745,0.6235,1.0000}

\begin{document}


\title{Multitude of singular jets during the collapse of drop-impact craters}

\author{Zi Qiang \surname{Yang}}
\affiliation{Division of Physical Science and Engineering,
King Abdullah University of Science and Technology (KAUST),
Thuwal, 23955-6900, Saudi Arabia}

\author{Yuan Si \surname{Tian}}
\affiliation{Division of Physical Science and Engineering,
King Abdullah University of Science and Technology (KAUST),
Thuwal, 23955-6900, Saudi Arabia}

\author{Sigur$\eth$ur T. \surname{Thoroddsen}}
\email{sigurdur.thoroddsen@kaust.edu.sa}
\affiliation{Division of Physical Science and Engineering,
King Abdullah University of Science and Technology (KAUST),
Thuwal, 23955-6900, Saudi Arabia}

\date{\today}

\begin{abstract}
\textcolor{black}{We study singular jets from the collapse of drop-impact craters, when the drop and pool are of different immiscible liquids.  These jets emerge from a dimple at the bottom of the rebounding crater, when no bubble is pinched off.   The parameter space is considerably more complex than for identical liquids, revealing intricate compound-dimple shapes.  In contrast to the universal capillary-inertial drop-regime, where the pinch-off neck radius scales as $R\sim t^{2/3}$, a purely inertial air-dimple has $R \sim t^{1/2}$ and is sensitive to initial and boundary conditions.  Capillary waves can therefore mold the dimple into different collapse shapes, with normalized jetting speeds one order of magnitude larger than for jets from bursting bubbles.  We study the cross-over between the two power-laws.  The fastest jets can pinch off a toroidal micro-bubble from the cusp at the base of the jet.}
\end{abstract}

\maketitle

\textcolor{black}{Singularities occur in many branches of physics from the gravitational collapse of a black hole \cite{chandrasekhar1992mathematical, choptuik1993universality} to the pinch-off of a drop from a faucet  
\cite{Eggers1997,brenner1997breakdown,eggers2015singularities,DayHinchLister1998}. 
The reduced length and time-scales near the singularity expose the important force balance governing the dynamics.}
\textcolor{black}{The pinch-off of a drop from a nozzle has a self-similar
conical shape with capillary-inertial scaling of the necking radius,
$R \sim t^{2/3}$ \cite{DayHinchLister1998}.  
In contrast the pinch-off of a bubble follows a purely inertial process with $R \sim t^{1/2}$ \cite{Taborek2005scaling,Eggers2007,Thoroddsen2007,schmidt2009memory}.
This modest difference in exponent values, hides a profound difference in the dynamical nature of the pinch-off.
For the purely inertia scaling the surface tension becomes irrelevant and there is strong dependence on initial or boundary conditions.  This memory of the boundaries has been demonstrated for the pinch-off of a bubble from an elliptic nozzle \cite{Lai2012curvature}.
For the collapse of impact craters \cite{Thoroddsen2018} showed that the finest {\it singular jets} emerge from the dimple collapse
with close to inertial scaling.}

\begin{figure}[b!]
  \centering
    \includegraphics[width=1.0\linewidth]{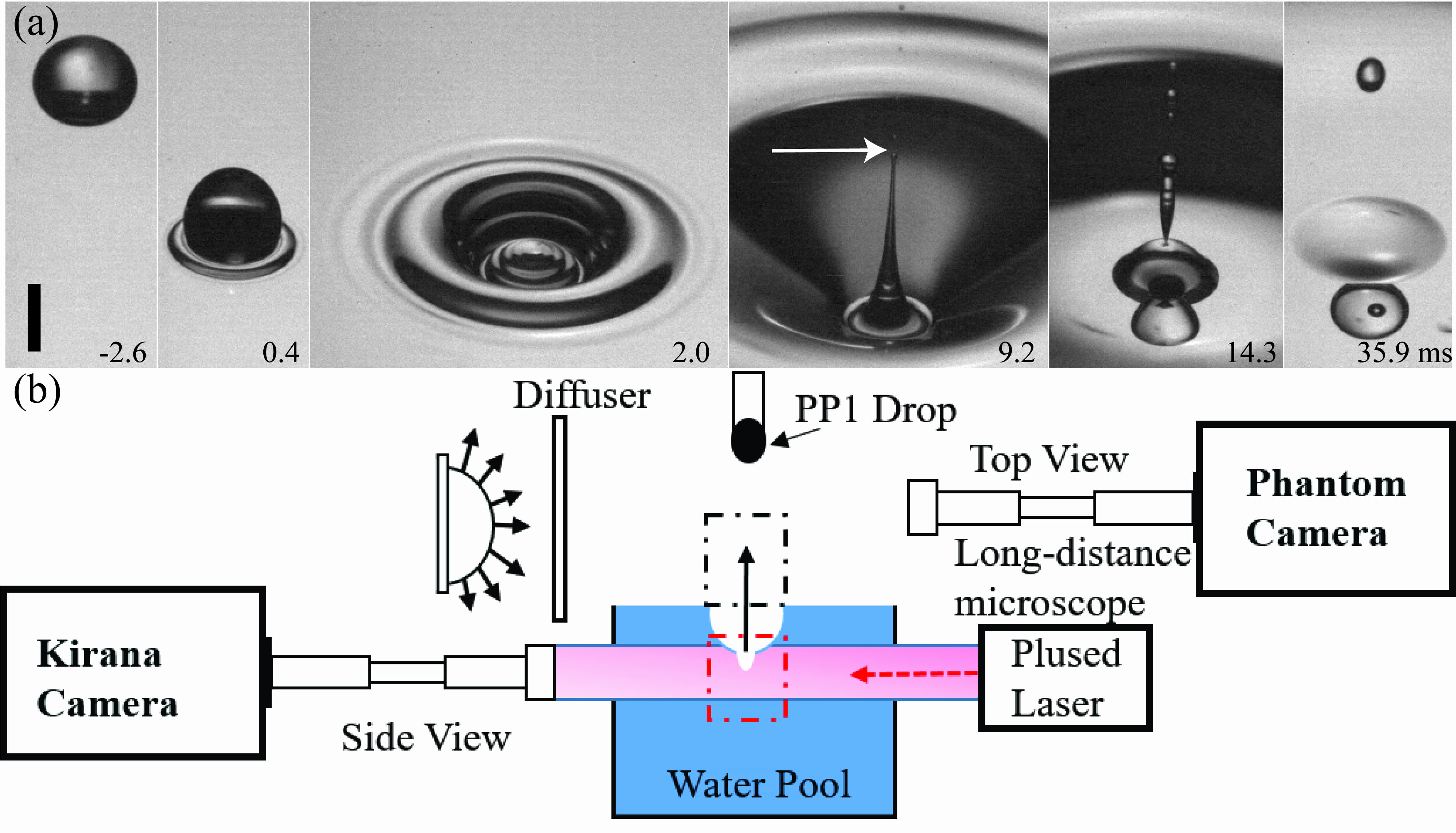}\vspace{-0.1in}\\ 
	\caption{(a) Video frames showing the typical impact crater collapse and jetting. The scale bar is 1mm long.  (b) Sketch of the experimental set-up, with two high-speed video cameras viewing from perpendicular directions.}
  \label{Fig_1}
  \vspace{-0.2in}
\end{figure}

\textcolor{black}{Fine jets can emerge from a free surface in numerous configurations, such as the oscillation of a free-falling drop pinched off from a nozzle \cite{Thoroddsen2007b}, 
from a bursting bubble at a pool surface \cite{Duchemin2002jet, Walls_2015, Lai2018, Deike2018}, 
drop impact on a superhydrophobic surface 
and from critical Faraday waves in vertically oscillated liquid layers \cite{longuet1983bubbles,zeff2000singularity}.
Herein we study jets forming by the collapse of drop-impact craters.
While numerous studies have looked at the crater collapse when the drop and pool are of the same liquid \cite{pumphrey1990entrainment,prosperetti1993impact,Liow2001splash,Thoroddsen2018}.
Few have studied a drop impacting a pool of a different immiscible liquid, 
mostly focusing on drop deformation \cite{fujimatsu2003interfacial} and fragmentation into smaller droplets \cite{lhuissier2013drop}.
We will show that the landscape for singular jetting becomes much more complicated in the immiscible case.} 
The overall setup is sketched in Fig. \ref{Fig_1}(b)
and is similar to that used in previous studies on this topic \cite{Thoroddsen2018}. 
The drop pinches off from a flat stainless steel nozzle and falls onto a pool surface 
contained in a square glass container (5 $\times$ 5 $\times$ 5 cm). 
The drop diameters are less than 2 mm, so capillary waves will not be reflected from the tank wall to influence the impact dynamics.
Our well-controlled experiments exhibited extreme sensitivity to boundary conditions,
as has been reported in \cite{Thoroddsen2018,Michon2017jet}.

\textcolor{black}{ 
Herein we use two immiscible liquids (Table 1).  The pool is purified water (Milli-Q), while the drop consists of {\it PP1} (Perfluorohexane, C$_6$F$_{14}$, from F2 Chemicals Ltd).  The PP1 is 1.71 times heavier than water and has very low surface tension $\sigma_d = 11.9$ mN/m.  The interfacial tension between water and PP1 is 48 mN/m.}

\textcolor{black}{Two high-speed cameras simultaneously observe the crater collapse and the jet rising above the pool surface.
The top camera (Phantom V2511) focuses on the jet above the liquid pool surface, 
while the other one studies the crater collapse below the liquid pool surface.  
The bottom Kirana camera can reach 5 Mfps at $\sim 1\; \mu$m/px resolution when using a 
long-distance microscope (Leica Z16 APO).  Back-lighting is produced by 350 W Sumita metal-halide lamp shone onto a diffuser,
or pulsed laser diodes (SI-LUX640, Specialized Imaging).}

\begin{figure*}[t]
  \centering
      \includegraphics[width=1.0\linewidth]{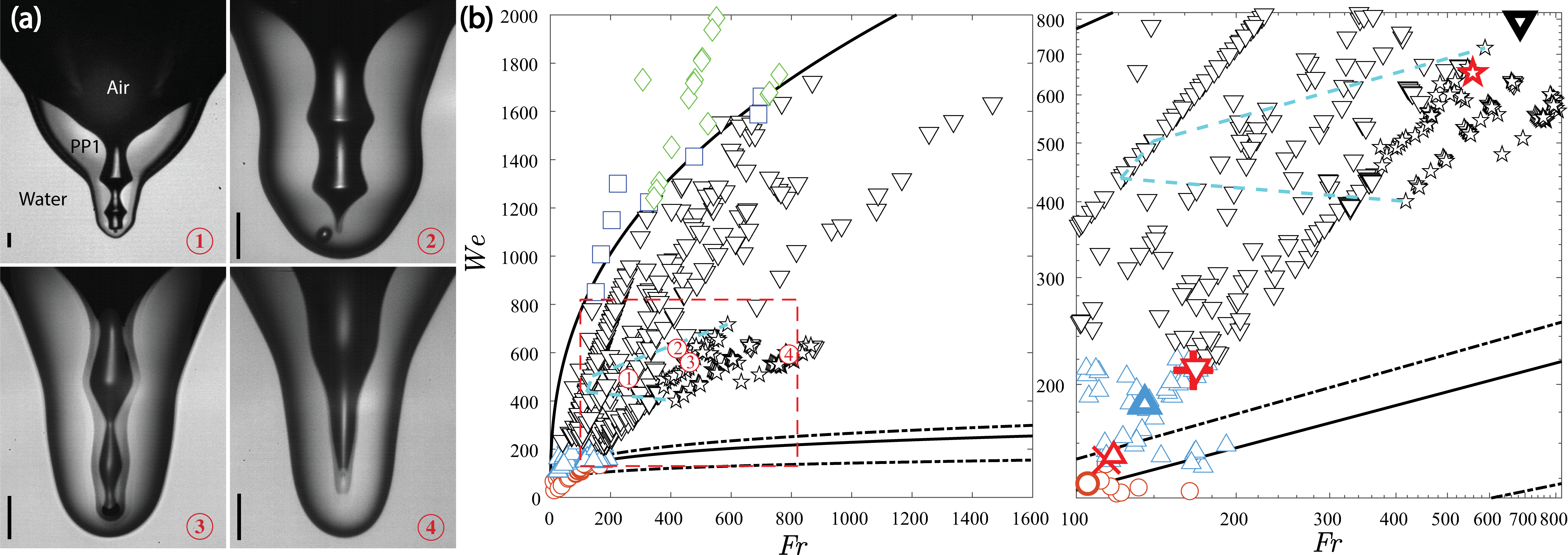}\vspace{-0.1in}\\ 
  \caption{(a) Typical dimple shapes for different impact conditions in the multi-dimple regime corresponding to the  circled red numbers in (b). Multi-pinch-offs dimple: \raisebox{.5pt}{\textcircled{\raisebox{-.9pt} {1}}}  
  $D=1.16$ mm, $U=1.7$ m/s, $Fr=259$, $We= 493$; 
  \raisebox{.5pt}{\textcircled{\raisebox{-.9pt} {2}}} $D=1.02$ mm, $U=2.1$ m/s, $Fr=421$, $We= 617$;
  \raisebox{.5pt}{\textcircled{\raisebox{-.9pt} {3}}} $D=0.93$ mm, $U=2.05$ m/s, $Fr=463$, $We= 560$ and singular telescopic dimple: \raisebox{.5pt}{\textcircled{\raisebox{-.9pt} {4}}} $D=0.73$ mm, $U=2.38$ m/s, $Fr=792$, $We= 593$.  The scale bars are 100 $\mu$m long.
  (b) Characterization of the dimples and jets in {\it Fr-We} space for drop impacts of immiscible liquids.
 The two dash curves are the bounds of the regular bubble entrapment measured by \cite{pumphrey1990entrainment,oguz1990bubble}. 
 The two solid curves mark the bubble entrapment region based on our study.
 (c) Enlarged region corresponding to the rectangular dashed box in (b). 
 The symbols correspond to different dimple shapes:
  (\textcolor{mypink1}{\textbigcircle}) no pinch-off shallow dimple; 
  (\textcolor{red}{$\times$}) first critical pinch-off (first singular jet) at the boundary between no and one bubble pinch-off;
  (\textcolor{red}{$\triangle$}) tiny bubble pinched off near first critical pinch off;
  (\textcolor{mypink2}{$\triangle$}) dimple pinch-off with bubble going out with jet; 
  (\textcolor{red}{$\plus$}) secondary critical pinch-off between bubble going out with jet and bubble entrapped in PP1 drop;
  (\textcolor{red}{$\triangledown$}) tiny bubble pinched off near secondary critical pinch-off;
  (\textcolor{red}{$\largewhitestar$}) singular telescopic dimple;
  (\textcolor{black}{$\triangledown$}) pinched-off bubble entrapped in PP1 drop;
  (\textcolor{blue}{$\square$}) liquid column break-up without dimple pinch-off; 
  (\textcolor{green}{$\Diamond$}) water entrapped in PP1 drop without pinch-off.
  The dashed cyan lines mark the region of multi-dimples.}
  \label{Fig_2}
\end{figure*}

We use a range of drop sizes $D$ = 0.60, 0.72, 0.85, 0.95, 1.2, 1.5 \& 2.0 mm 
and by varying the drop release-height we produce impact velocities between 0.1 to 3.9 m/s.
The corresponding range of Reynolds, Weber and Froude numbers, based on the drop liquid properties are: 
$Re = \rho_d DU/\mu_d = 374 - 10,200$;
$We = \rho_d DU^2 /\sigma_d  = 10 - 2,000$; $Fr = U^2/(gD) = 10 - 1,500;$
where $g$ is gravity, $\rho_d$ and $\mu_d$ drop density and dynamic viscosity.
\begin{table}[b]
\caption{\label{tab:table1}%
Liquid properties.}
\begin{ruledtabular}
\begin{tabular}{cccccc}
\textrm{Liquid }&
\textrm{Density  }&
\textrm{Viscosity  }&
\textrm{Surface   }&
\textrm{Capillary  }&
\textrm{Refract  }\\
\textrm{ }&
\textrm{ }&
\textrm{ }&
\textrm{ tension}&
\textrm{ length}&
\textrm{ index}\\
\textrm{ }&
\textrm{ $\rho$ }&
\textrm{ $\mu$ }&
\textrm{ $\sigma$  }&
\textrm{ $L_c$ }&
\textrm{ $n$}\\
\textrm{ }&
\textrm{  $[g/cm^3]$}&
\textrm{ $[mPa.s]$}&
\textrm{  $[mN/m]$ }&
\textrm{ $[mm]$}&
\textrm{ $n$}\\ \hline
PP1 & 1.71& 0.81& 11.9& 0.84& 1.25 \\
Water & 0.996& 1.004& 72.1& 2.72& 1.33 \\
\end{tabular}
\end{ruledtabular}
\end{table}


\begin{figure}[!hbp]
  \centering
	  \includegraphics[width=0.76\linewidth]{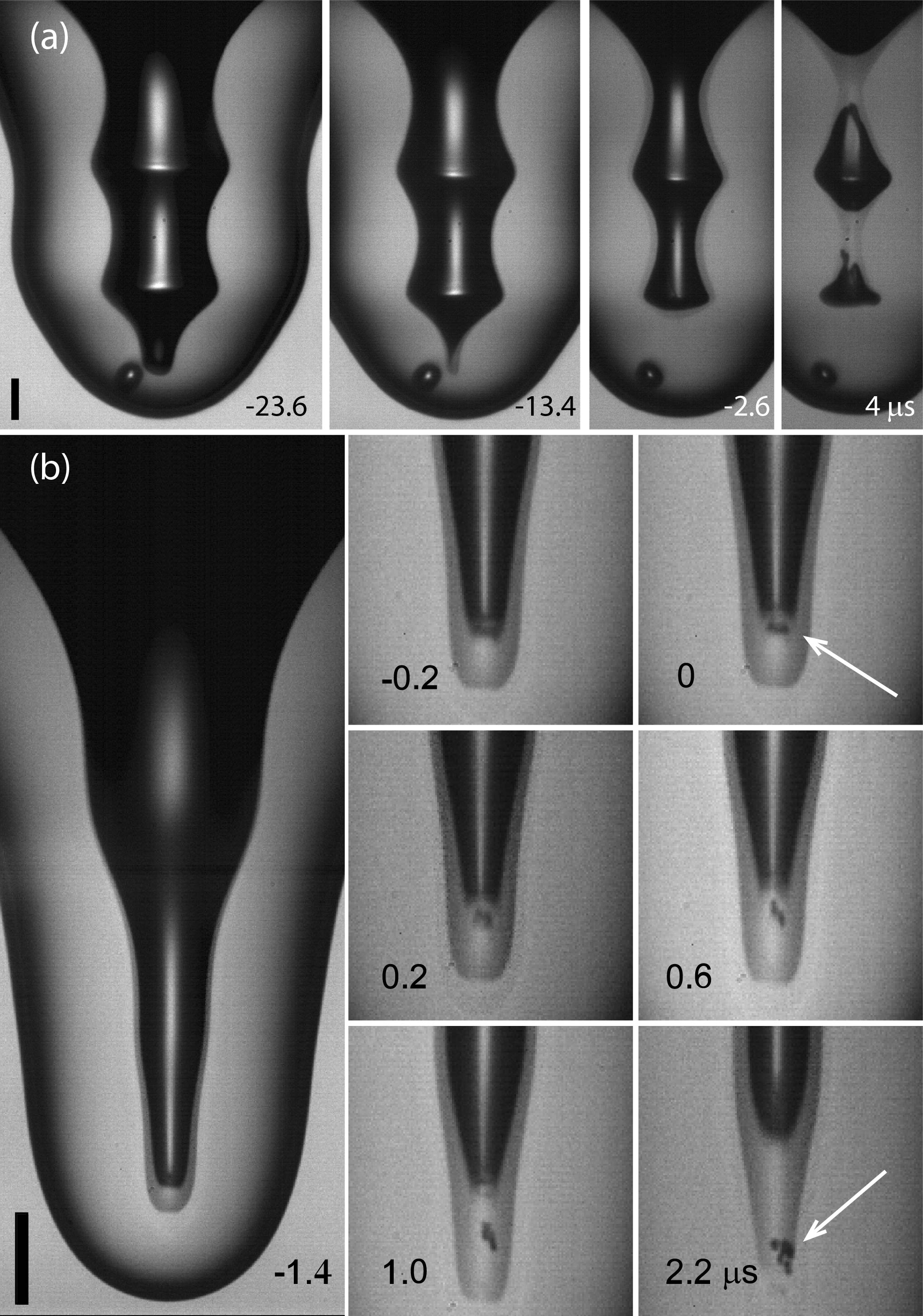}\vspace{-0.1in}\\
  \caption{(a) Multi-pinch-offs dimple shape, corresponding to \raisebox{.5pt}{\textcircled{\raisebox{-.9pt} {2}}} in Figure \ref{Fig_2}(a);
  (b) Micro-bubble shedding from the cusp at the base of the singular jet, for $D=0.82$ mm, $U=2.21$ m/s, $We= 609$, $Fr=569$. The arrows point at the shed micro-bubbles. The image-sensor has strong ghosting from every 10th frame. The scale bars are 50 $\mu$m long.}
  \label{Fig_100} 
\end{figure}


\begin{figure*}[!ht]
  \centering
	  \includegraphics[width=0.9\linewidth]{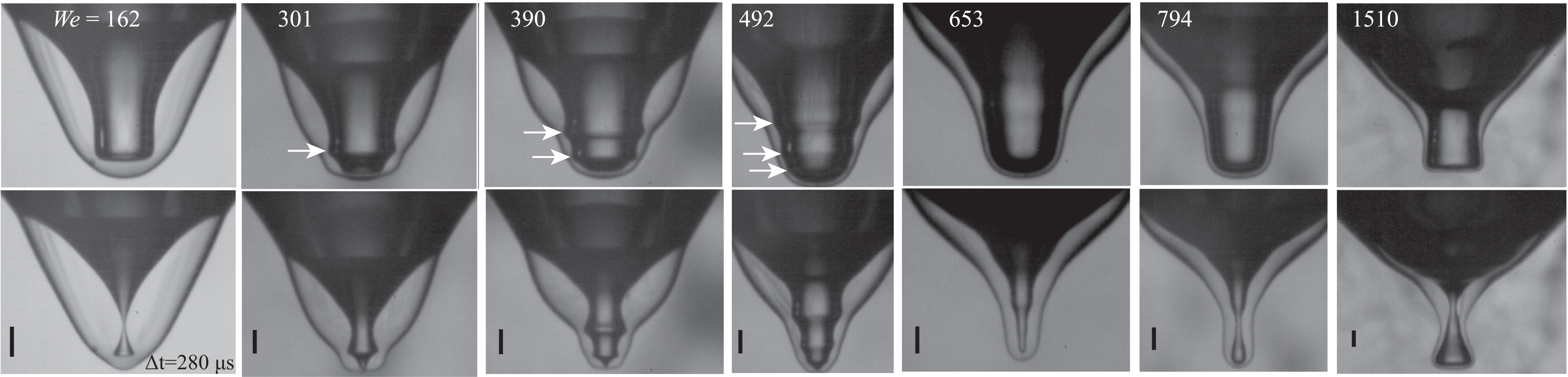}\vspace{-0.1in}\\
  \caption{Capillary wave-shapes on the dimple for a range of $We$, 
  for $D=0.935\pm 0.025$ mm and impact velocity increasing from left to right:
  $U=$ 1.09, 1.48, 1.72, 1.91, 2.23, 2.47 \& 3.37 m/s.
The bottom row is shown at the most singular point during the collapse,
with the top row 280 $\mu$s earlier.  The arrows point out capillary wave-crests. 
The scale bars are 200 $\mu$m long.}
  \label{Fig_4}
\end{figure*}

\begin{figure*}
  \centering
	  \includegraphics[width=0.9\linewidth]{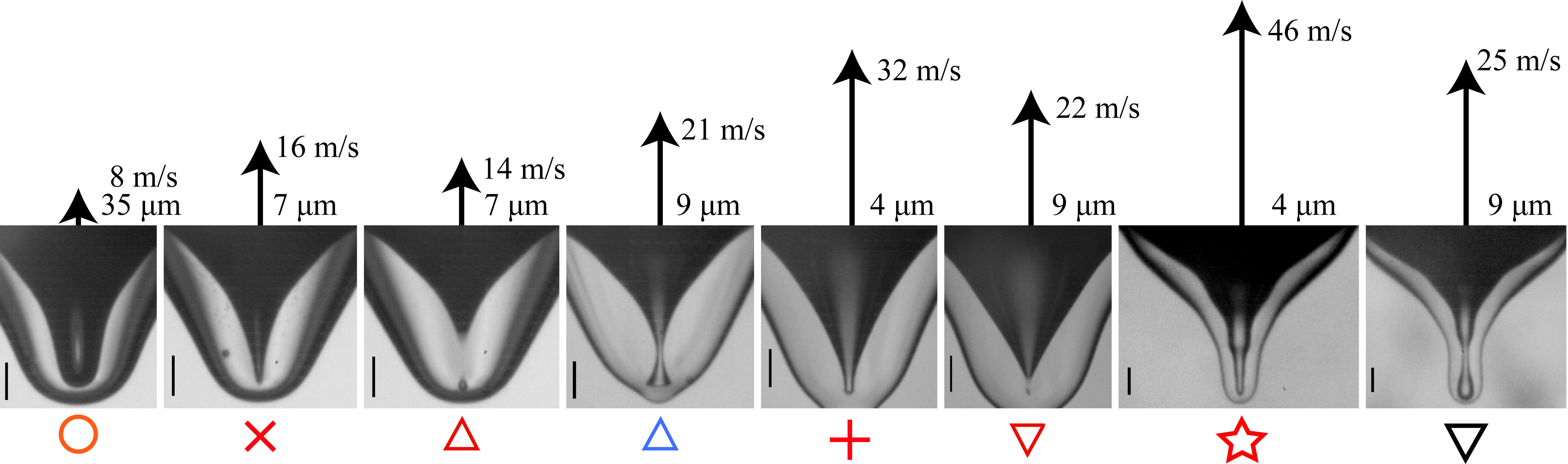}\vspace{-0.1in}\\
  \caption{Overview of dimple shape and jet velocity versus $We$, for drop size 0.92 mm.
  The arrow lengths indicate the jet velocities. 
  The Weber number grows from left to right ($We=137, 139, 153, 186, 211, 213, 653, 794$).
  The scale bars are 200 $\mu$m.} 
  \label{Fig_6}
\end{figure*}

\textcolor{black}{
The impact forms a hemispheric crater into the pool surface, 
with the drop liquid stretched out into a thin continuous layer coating it.
The subsequent rebound can form a bottom dimple whose collapse produces singular jets \cite{Rein1996transitional, Michon2017jet, Thoroddsen2018}.
The free surface of this dimple therefore remains between air and the PP1 drop liquid.
Figure \ref{Fig_2} shows the regime where a dimple forms at the bottom of the crater, during its collapse.  
This occurs at much larger $We$ (based on drop properties), than for 
the classical regime (dashed lines) where the dimple entraps a bubble 
for identical liquids in both drop and pool \cite{pumphrey1990entrainment,prosperetti1993impact}.}

\textcolor{black}{Figure \ref{Fig_2}(a) shows a prominent new feature of the dimples, 
i.e. capillary waves travelling down towards their tips.  
Some of these shapes evolve multiple pinch-offs, like the two shown in Fig. \ref{Fig_100}(a).
Figure \ref{Fig_4} shows the evolution of the wave-shapes along a cut through parameter space,
where we keep the drop size fixed
while increasing the impact velocity,
to span a range of $We$ from 162 to 1510.  The number of visible wave crests grows from one to three (middle panels) 
and then the dimple column becomes smooth again (last panels).
The lower row shows the corresponding final pinch-off shape, which includes a {\it singular telescopic dimple},
where no pinch-off occurs, but the fastest jets are ejected out of the crater ($We=653$).
This intriguing telescopic shape occurs in a very limit regime, 
within the more common {\it multi-pinch-offs} shown in three of the panels in Fig. \ref{Fig_2}(a).}

From these realizations it becomes clear that the classical picture of singular jets only appearing at the boundaries of the regular bubble-entrapment regime, 
no longer applies and the phase of these capillary waves can induce singular jets at more $We$ values.  
This is shown for $D=0.92$ mm in Fig. \ref{Fig_6}, with the corresponding jet velocities.  
Here there are three separate $We$ values where no bubble is pinched off and a fast jet is produced
(panels 2, 5 \& 7).  
See Supplementary Fig. S1 for the shape and breakup of these jets as they emerge from the crater \cite{Yarin1993}.
The fastest and thinnest jet is observed for the {\it telescopic-dimple} case ($We=653$), which corresponds to the narrowest angular span of the air cylinder,
where maximum flow volume can be focused into the base of the jet.
This narrow shape looks reminiscent of the capillary-driven retraction of a conical drop,
studied by Brasz {\it et al.} \cite{Bird2018}.

\textcolor{black}{This shows that not only can the boundary conditions break the axisymmetry of the collapse of a pinching air cylinder \cite{Taborek2005scaling,keim2006breakup,Lai2012curvature}, 
but they can also imprint a large variety of axial shapes on the free surface of the dimple,
thereby modifying its singular collapse.}

\begin{figure} 
  \centering
     \includegraphics[width=1.0\linewidth]{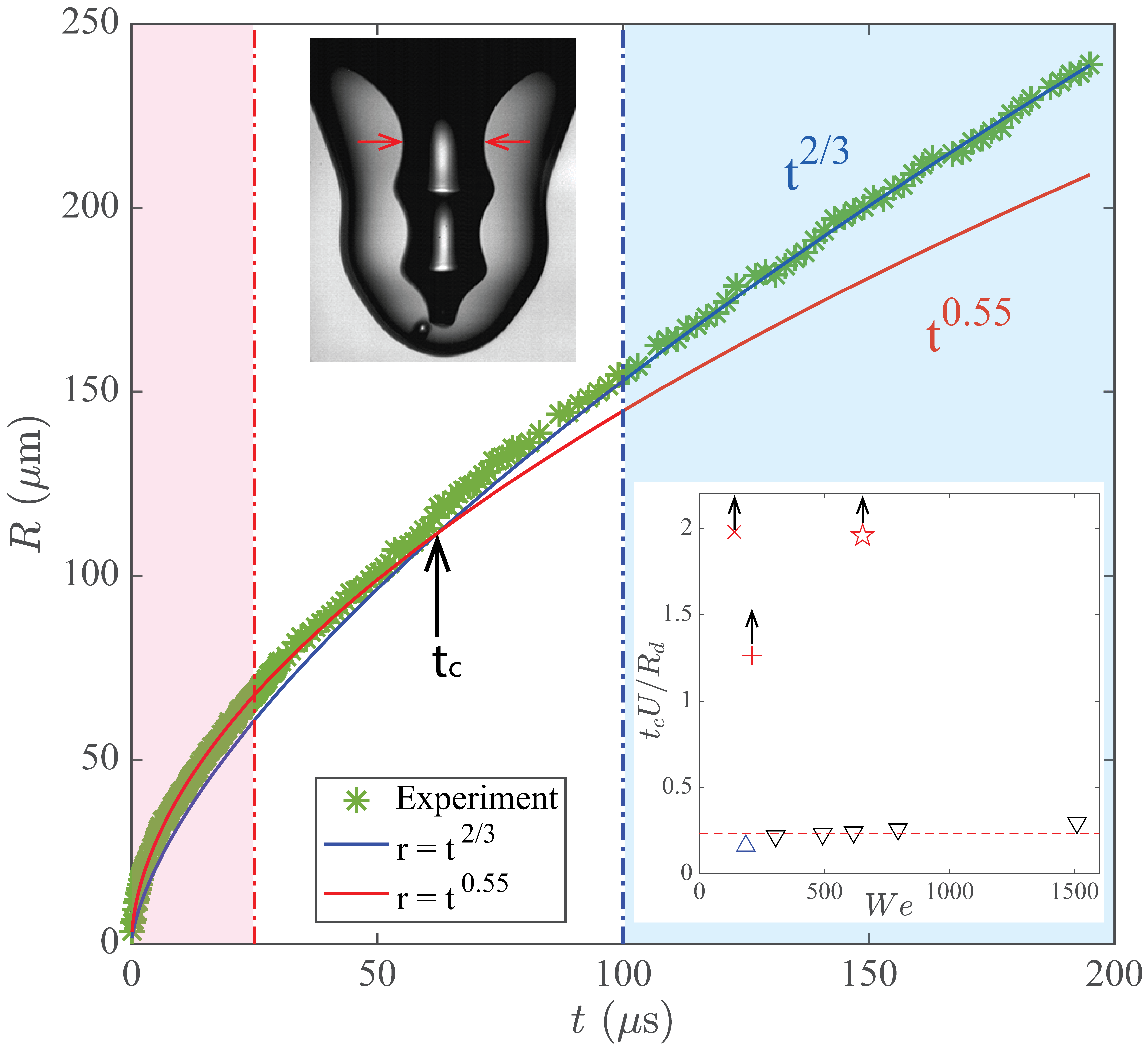}\vspace{-0.15in}\\
	\caption{Scaling of the dimple radius vs time before pinch-off. 
	There is a transition of power-law exponents from 2/3 to 0.55 closest to the pinch-off.
	The background shading marks the validity of each, with the arrow indicating the approximate cross-over time $t_c$.  
	The data is taken from two video clips spanning time-scales from 100 ns to 200 $\mu$s before pinch-off.  
	The corresponding log-log-plots are included in Suppl. Materials.
	The inset shows how $t_c$ normalized by the impact time $D/U$ changes with $We$, 
	for dimple pinch-off (\textcolor{mypink2}{$\triangle$} \& $\triangledown$) and singular jets (\textcolor{red}{$\times$}, \textcolor{red}{$\plus$} \& \textcolor{red}{$\largewhitestar$}).
	The vertical arrows indicate these are lower bounds.}
  \label{Fig_5}
\end{figure}

\textcolor{black}{What is the role of capillary waves in setting up the dimple for the inertial focusing?
For the singular jets the dimple dynamics have until recently been formulated in the self-similar capillary-inertial formalism \cite{zeff2000singularity,Duchemin2002jet,Deike2018}, 
while the final cylindrical collapse has been shown to follow pure inertial focusing \cite{Thoroddsen2018,Gordillo2019}.
One can therefore expect a dynamical transition in the vicinity of the final jet formation.
In Figure \ref{Fig_5} we track the radius of pinch-off neck for a typical dimple, shown in the inset.
There is a clear cross-over in the nature of the dynamics from capillary-inertial $R\sim t^{2/3}$ 
to purely inertial with $R\sim t^{0.55}$ at $t_c \simeq 65\; \mu$s before pinch-off, as marked by the arrow.
The inset shows that the cross-over time scales with the impact time $t_c \simeq 0.235 R_d/U$ for the pinch-off cases.
On the other hand, for singular jetting the cross-over time occurs much earlier, 
irrespective of $We$.}

Herein, we report a plethora of new dimple shapes, which occur following a drop impact on an immiscible pool.  This includes multiple pinch-offs and many discrete $We$ where singular jetting is observed.
Questions remain:  what determines the minimum diameter of the singular dimple
and thereby its maximum velocity?
The smallest singular dimple width is here $\simeq 12\; \mu$m which is similar to the 15 $\mu$m observed by Thoroddsen {\it et al.} \cite{Thoroddsen2018},
who used a liquid which is an order of magnitude more viscous.  
This suggests viscous cut-off is not at play for the much lower viscosity of our PP1 drop.  
We can speculate that cavitation or vortex-shedding instability \cite{Thoraval2012} in the cusp at the base of the jet prevents smaller jet sizes,
as we see by the micro-bubbles shed at the base of the jet in Fig. \ref{Fig_100}(b).
The expansion of the bubble-volume in the last panel indicates the large localized pressure driving up the singular jet \cite{tran2016hydrodynamic,Gordillo2019,Thoroddsen2018}.
\textcolor{black}{We conclude that our singular jets differ from bubble-bursting jets,
in fundamental ways.  First, the dimple shapes are not self-similar during the collapse \cite{Duchemin2002jet,Lai2018}.
Secondly, Figures \ref{Fig_4} \& \ref{Fig_6} show clearly that the Ohnesorge number, which is approximately constant, 
$Oh=\mu_d/\sqrt{\rho_d R_c \sigma_d}\simeq 0.0054$, based on the maximum crater radius $R_c=1.1$ mm, is not sufficient to describe the dynamics, 
as is suggested for bursting-bubble jetting \cite{Ganan-Calvo2017,Ganan-Calvo2018,Gordillo2019}.
It is a clear indication of the extreme focusing of energy that the maximum jetting velocity
$v_j = 46$ m/s is $\sim 580 \pm 30$ times the capillary velocity $v_{\sigma}=\sqrt{\sigma_d/(\rho_d R_{c})}$. 
This is an order of magnitude faster than predicted for the bursting bubbles \cite{Deike2018,Ganan-Calvo2018,Gordillo2019}.
The jet diameters of 4 $\mu$m are also two orders of magnitude thinner than those predicted for the bursting bubbles \cite{Ganan-Calvo2018},
see also \cite{Gordillo2018}.}
\textcolor{black}{Finally, we point out that while the final inertial focusing occurs on tens of $\mu$m lengthscale, the larger-scale liquid inertia is here a function of time, 
owing to the local thickness of the drop liquid around the dimple,
which becomes thinner with increasing $We$.
This effect can be investigated by changing the relative density of the two liquids, in future experiments.}
\bibliographystyle{apsrev4-2} 
\bibliography{ms} 

\end{document}